\documentclass[letterpaper,10pt,conference]{IEEEtran}
\IEEEoverridecommandlockouts
\makeatletter
\def\ps@headings{%
\def\@oddhead{\mbox{}\scriptsize\rightmark \hfil \thepage}%
\def\@evenhead{\scriptsize\thepage \hfil \leftmark\mbox{}}%
\def\@oddfoot{}%
\def\@evenfoot{}}
\makeatother
\usepackage[utf8]{inputenc} 
\usepackage[T1]{fontenc}
\usepackage[cmex10]{amsmath}
\usepackage{amsfonts,ulem}
\usepackage{algorithm,amssymb,url}
\usepackage{graphicx,color,citesort} 
\usepackage{algorithmic}
\usepackage{cite}
\usepackage{epsfig}
\usepackage{tabularx}
\usepackage{bbm}

\newtheorem{proposition}{Proposition}

\newtheorem{theorem}{Theorem}

\ifCLASSOPTIONcompsoc
\usepackage[caption=false,font=normalsize,labelfont=sf,textfont=sf]{subfig}
\else
\usepackage[caption=true,font=footnotesize]{subfig}
\fi

\begin{document}
\title{Self-Optimizing Mechanisms for EMF Reduction in Heterogeneous Networks}
\author{Habib~B.A.~Sidi, Zwi~Altman and Abdoulaye~Tall\\\\
Orange Labs R\&D, 38/40 rue du General Leclerc,92794 Issy-les-Moulineaux\\
Email: \{habib.sidi,zwi.altman,abdoulaye.tall\}@orange.com\\
}


\maketitle

\begin{abstract}
This paper focuses on the exposure to Radio Frequency (RF) Electromagnetic Fields (EMF) and on optimization methods to reduce it. Within the FP7 LEXNET project, an Exposure Index (EI) has been defined that aggregates the essential components that impact exposure to EMF. The EI includes, among other, downlink (DL) exposure induced by the base stations (BSs) and access points, the uplink (UL) exposure induced by the devices in communication, and the corresponding exposure time. Motivated by the EI definition, this paper develops stochastic approximation based self-optimizing algorithm that dynamically adapts the network to reduce the EI in a heterogeneous network with macro- and small cells. It is argued that the increase of the small cells’ coverage can, to a certain extent, reduce the EI, but above a certain limit, will deteriorate DL QoS. A load balancing algorithm is formulated that adapts the small cell’ coverage based on UL loads and a DL QoS indicator. The proof of convergence of the algorithm is provided and its performance in terms of EI reduction is illustrated through extensive numerical simulations.
\end{abstract}

\IEEEpeerreviewmaketitle

\section{Introduction}
The reduction of exposure to EMF is one of the challenging problems in radio access technology (RAN) that attracts the attention of different stakeholders, from the general public, to telecoms industry and regulatory bodies. The problem has become even more prominent in face of the frantic race toward the increase of network capacity and optimizing its performance and quality of service (QoS) \cite{shuping11}. 
In fact the introduction of multiple new technologies in the network goes along with the deployment of several new radiating antenna and transmission towers. Up today, the assessment of RF-EMF exposure has been focused separately on the exposure induced by personal devices on the one hand, and that of the network equipment such as base stations (BSs) and access points on the other hand. In the framework of the FP7 European research project LEXNET, an exposure metric has been proposed, denoted as the Exposure Index (EI). The EI aggregates the exposure from both personal devices and that from BSs and access points, giving rise to a single and more realistic network parameter for exposure. It reflects the contribution to exposure from different technologies and, among others, takes into account the exposure duration from the different radiating sources, the utilized frequencies, environment, services etc. It is argued that the EI could be used in developing deployment strategies for RANs, optimization, self-optimization and management techniques for reducing EMF, using a more realistic metric. 
 Recently the issue of reducing the overall and individual level of exposure has been addressed and becomes one of the main concerns from the perspective of the network users. In \cite{foster2007radiofrequency} for example, a measurement of RF field from WiFi access points against a background of RF fields in the environment over the frequency range 75 MHz-3 GHz is performed to quantify the exposure that a bystander might receive from the laptop. In \cite{emfexpofrompervasivecomp} the different factors influencing RF exposure in mobile networks are treated in a systematic manner for most relevant wireless standards relying on their RF characteristics. The most relevant levers for limiting future exposure levels are presented.

Nevertheless the risk perception linked to EMF exposure from the network users is different. \cite{riskRFD22lexnet} illustrates the biased view on RF exposure of network users, who mainly focus on the radiating antennas of telecommunication towers while ignoring or giving little importance to radiation from user equipment close to the body. However the two sources of radiation are strongly correlated. Leading works in the area of EMF exposure reduction consider solutions for uplink (UL) and downlink (DL) transmissions separately. In compliance with the definition of the EI, we address the combined effect of radiation from UL and DL transmissions. 

The purpose of this work is to propose a novel Self Organizing Network (SON) approach for reducing the overall EMF exposure, expressed in terms of the EI. The EI depends on highly complex set of network data and parameters that are too complex to be handled instantly. For this reason, we adopt the strategy of self-optimizing intermediate Key Performance Indicators (KPIs) that impact the EI of both UL and DL transmissions.
Once transmit (Tx) and Received (Rx) powers are calculated, EMF exposure can be evaluated using transformation tables (evaluated in \cite{EID24lexnet} using measurements and electromagnetic simulators), and the corresponding EI. 
The proposed SON algorithm is a load balancing algorithm, based on a stochastic approximations. It adapts the small cells' coverage based on UL loads and on DL QoS indicators. The rationale for the proposed solution is that, to a certain extent, by off-loading macro-cell traffic towards small cells, UL transmission of cell edge users is decreased. However, above a certain cell range extension, DL QoS can be jeopardized, and should be therefore included in the SON algorithm.

The contributions of this paper are the following:
\begin{itemize}
\item[\it (i)] A novel control stochastic load balancing algorithm based on recursive inclusion, which takes into account UL loads and DL QoS constraints. 
\item[\it (ii)] The proof of convergence of the proposed algorithm is provided referring to the developments on stochastic differential inclusions in \cite{borkar2008stochastic}.
\item[\it (iii)] Performance analysis through flow level simulations of the designed mechanism is provided.
\end{itemize}
The paper is organized as follows: Section~\ref{sec:model} gives the description of the network settings, the problem formulation and the methodology used in the paper. In Section~\ref{sec:flowlevelmetric} we present the different metrics and flow level KPIs. Section~\ref{sec:sondev} develops the proposed load-balancing approach to reduce EMF exposure. We evaluate numerically the performances obtained upon activation of the SON mechanism in Section~\ref{sec:numanalysis} and discuss the results. Section~\ref{sec:conclusion} eventually concludes the paper.
 

\section{Model}\label{sec:model}
\subsection{Model description and formulation}
Consider a heterogeneous network(HetNet) deployment with several operating macro- and small cells (SCs) located close to the edge of each macro-cell coverage area. We assume that the SCs can be activated whenever additional capacity is needed to serve the traffic in the cell. 
All nodes (macro and SCs) use the same frequency bandwidth. Information can be exchanged between the macro- and small cells in their coverage area using physical or logical links such as the X2 interface in LTE. The system considered in this work matches with LTE network requirements for both UL and DL transmissions. More precisely we are focused on OFDMA based transmissions in UL and DL, although it is noted that the proposed methodology can be adapted to other radio access technologies. 

\begin{figure}[!t]
\centering
\includegraphics[width=3.5in]{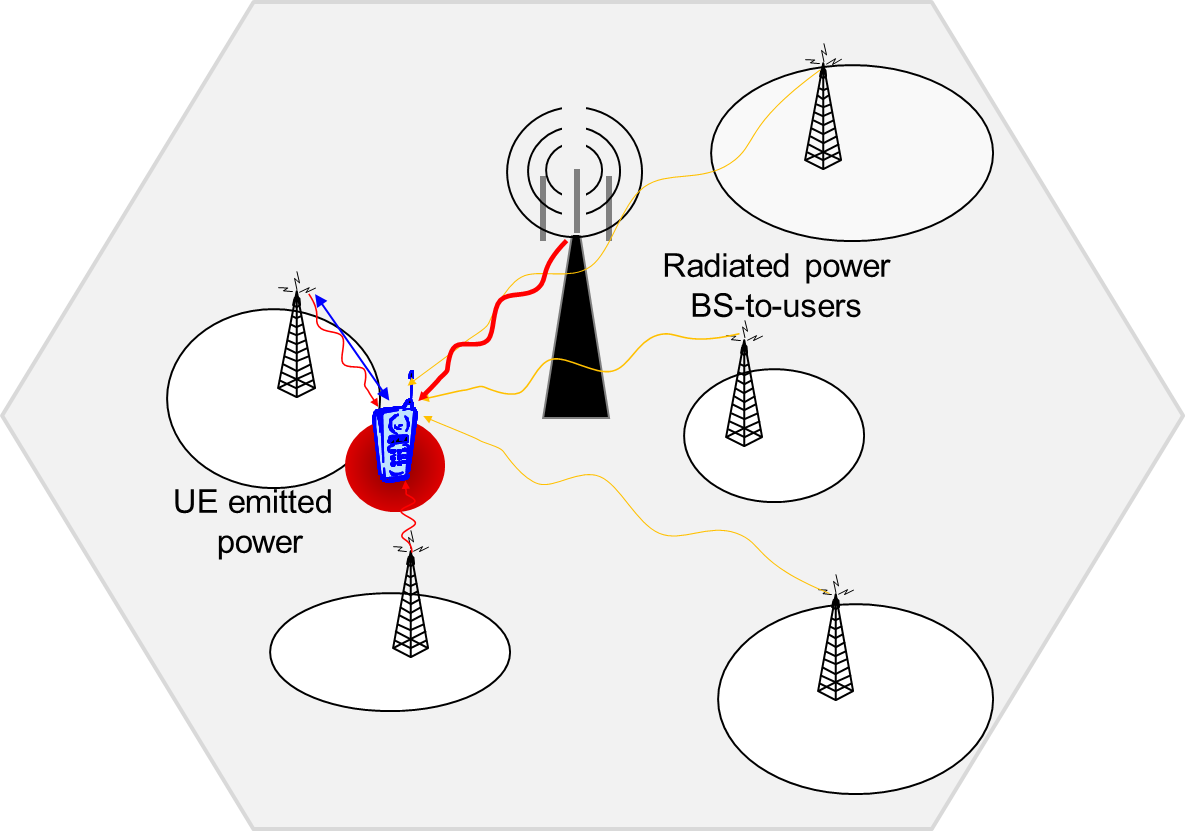}
\caption{EMF sources in HetNets}
\label{netexposure}
\vspace{-0.5cm}
\end{figure}

In radio access networks, EMF exposure comprises two components \cite{EID24lexnet}: UL transmissions from user equipment (UE) and DL transmissions from all the BSs in the network (see Figure~\ref{netexposure}). 
In order to  reduce the EI in the network one can thus seek to minimize a well defined cost function that combines the joint effect of both UL and DL transmissions. More specifically we focus on reducing the average level of UL Tx power from UEs to their serving cells as an intermediate metric to reduce EI. Such objective can be achieved by increasing SCs' coverage, and off-loading macro-cell traffic towards SCs. 
In fact, off-loading the macro-cell with low power nodes such as SCs allows, not only to bring more users to transmit to a closer serving cell with reduced power, but also to a certain extent, to increase the network capacity by off-loading loaded nodes \cite{combes2012self}. However, as more users are off-loaded to the SCs it is likely that we observe a fast decrease of the UL/DL QoS. This is due to a large number of new-interferers (in UL) inside the macro-cell coverage area and the additional interference produced by the SCs that see their loads increase (in DL). 
We propose to formulate the EI reduction problem as a QoS constrained optimization problem which we address using an off-loading method relying on SCs  coverages' expansion/contraction. Indeed by expanding their coverage, SCs can collect more users from the macro-cell relying on the user-to-cell best server attachment criterion \cite{3gpp.36.304}. It is noted that coverage extension is achieved not by increasing the Tx power of the SCs but rather by increasing/decreasing the value of the Cell Individual Offset (CIO) used in the network selection/re-selection and handover (HO) procedures \cite{3gpp.36.331,3gpp.36.304}. Specifically, during idle mode or HO procedures, the mobile compares through a set of measurements the DL received power plus $CIO$ values from each of its neighboring cells. 
This is the case for Event A3 measurement report triggering \cite{3gpp.36.331}. Then, a ranking of all the available cells including the serving cell is done using the offset ratio of the candidate cells to the serving cell. Typically, the selection of cell $n$ or HO when user is attached to cell $s$ is made if:
\begin{equation} Q_{meas,s} + Q_{Hyst} < Q_{meas,n} - Q_{offset_{s,n}},
\end{equation}
with  $Q_{meas,s}$, $Q_{meas,n}$ being the measured averaged Reference Signal Received Power (RSRPs), $Q_{Hyst}$ is the hysteresis value and $Q_{offset_{s,n}} = CIO_s - CIO_n$ for intra-frequency cell selection. In our particular case, the mobile compares the received signal power plus $CIO$ values from the SCs to the received signal power from the macro-cell to define its attachment. The problem thus formulates as follows:
\begin{eqnarray}\label{optimpbm}
\underset{{\bold P}}{min}\; (\bar{\rho}_M^{UL} - \bar{\rho}_{SC}^{UL}),\quad& \forall \;SCs, \\ \nonumber
 s.t. \; QoS_{DL}>QoS_{target}&
\end{eqnarray}
where $\bar{\rho}_M^{UL}$ is the UL load of the macro-cell, $\bar{\rho}_{SC}^{UL}$ the UL load of each small cell in the network, ${\bold P}$ is the optimization vector variable, with element ${\bold P_s}$ the DL pilot power plus $CIO$ of cell s, $QoS_{DL},\;QoS_{target}$ are respectively the actual and target DL QoS levels.
The problem is combinatorial and depending on the network dynamics a potential solution can stuck at the bounds of the constraint set. We propose in Section \ref{sec:sondev} an iterative load-balancing solution to (\ref{optimpbm}) based on stochastic approximation.

\subsection{Network model}
This section presents the network model and the assumptions  herein considered. Let $N$ be the number of BSs (macro and SCs) within a bounded area $\mathcal{A}\in \mathbb{R}^2$. We also denote by $n_s$ the number of users in cell $s$.
\subsubsection{Uplink}
We denote by $h_{r,s}$ the pathloss (including antennas gain and shadowing) between a location $r \in \mathcal{A}$ and a given BS $s$. Transmissions occur in the UL and we assume that the system operates on a bandwidth $W=N_{PRB}*W_{PRB}$, with $N_{PRB}$ the number of available resource blocks per BS and $W_{PRB}$ the bandwidth in $Hz$ of each PRB. A UE at location $r$ is assigned a number $M$ of resource blocks and transmits with a total  power of $P_{r,s} = \sum_k P^{(k)}_{r,s}$ to BS $s$ where $P^{(k)}_{r,s}$ is the Tx power on resource $k$ and $\mathcal{P} = (P_{r,s}^{(k)})_{\small\hspace{-0.1cm}\begin{array}{c}r\in{\mathcal{A}},1<s<N,\;0<k<N_{PRB}\end{array}\normalsize}$ is the Tx power matrix on the UL. 

Note that we do not make here any assumption on the way the total Tx power of the mobile is split over the allocated sub-bands. It is discussed in  \cite{kkimoptpowalloc05} and \cite{JLimeqpowalloc06} that the achieved gains by subdividing the total power equally over allocated resources is negligible compared to higher complexity of optimal allocation.
In the following we thus consider equal distribution of the total Tx power over the allocated resources such that $P^{(k)}_{r,s} = \frac{P_{r,s}}{M}\; \forall k$.
A closed-form expression of $P_{r,s}$ is given by \cite{lteuplinkFPCnsn08}: \small
\begin{equation}  
P_{r,s}=min\{P_{max}, P_0+10\log_{10} M+\alpha\cdot h_{s,r}+\Delta_{mcs} + f(\Delta_r)\}, \label{ultrpower}
\end{equation}
\normalsize
where $P_{max}$ is the maximum Tx power of the UE, $P_0$ is a UE or a cell-specific parameter, $\alpha$ is the cell specific pathloss compensation factor, $h_{s,r}$ is the DL pathloss measured at the UE, $\Delta_{mcs}$ is specified at the UE by the upper-layers and $\Delta_{r}$ is a UE-specific closed-loop correction value with a relative or absolute increase depending on the function $f()$.
A communication is possible between user in $r$ and a BS $s$ whenever she is in the coverage area $\mathcal{A}_s\subset \mathcal{A}$  of $s$ defined by a best server attachment criteria: 
\begin{equation}\label{covarea}
\mathcal{A}_s = \{r\in \mathcal{A}\;|\; s= arg\max_s(h_{s,r}(P_s^{(0)}+CIO_s))\},\end{equation} 
where $P_s^{(0)}$ is defined as the DL pilot power of the cell. Using this notation, we can write the signal to interference plus noise ratio ($SINR$) for UL transmission, $SU^{(k)}_r$ of resource $k$ for user in location $r$ as: 
\begin{equation}\label{sinrlwbound}
SU^{(k)}_r = \frac{P^{(k)}_{r,s}h_{r,s}}{W_{PRB}N_0 + \underset{r'\in \mathcal{A}_{s'},s'\neq s}{\sum} P^{(k)}_{r',s'}h_{r',s}},
\end{equation}
with $N_0$ being the thermal noise spectral power density. Here we consider that $P^{(k)}_{r',s'}=0$ when user in $r'$ from station $s'$ does not use resource $k$. Note that $SU^{(k)}_r$ is a function of $\mathcal{P}$ which is omitted in the expression for simplicity. It is then possible to derive the UL spectral efficiency of user in location $r$ as a function of the $SINR$, $\phi(SU^{(k)}_r) \leq \log_2(1 + SU^{(k)}_r)$ bounded by the maximum Shannon capacity. 
The corresponding rate follows up as: 
\begin{equation}\label{eq:rateup}
R_U(r,\mathcal{P}) = \sum_{k=1}^{N_{PRB}}W_{PRB}\int^{+\infty}_0 \phi(SU^{(k)}_r\, x) p_{\xi}(x)dx,
\end{equation}
where $p_{\xi}(x)$ is the probability density function of the ergodic channel fading process that is averaged over each $PRB$. It is noted that in (\ref{eq:rateup}) the user is allocated all the available PRBs and hence achieves its peak rate. We assume here a Round Robin scheduler, that allocates ($\frac{N_{PRB}}{n_s}$) PRBs when $n_s$ mobiles are present in cell $s$. The expression of (\ref{ultrpower}) becomes: 
\begin{equation} 
P_{r,s} = min\{P_{max},  P_0 + 10\log_{10} (\frac{N_{PRB}}{n_s}) + \alpha \cdot h_{s,r}\} 
\end{equation}
\subsubsection{Downlink}
The same analysis is done for the DL and we derive the DL $SINR$ and users rates for each cell $s$: 
\begin{equation}\label{sinrlwbound2}
SD^{(k)}_r= \frac{P^{(k)}_{s}h_{s,r}}{W_{PRB}N_0 + \underset{s'\neq s}{\sum} P^{(k)}_{s'}h_{s',r}}
\end{equation}
where $P^{(k)}_{s}$ is the data channel power applied by BS $s$ on resource $k$ and $h_{s,r}$ is the channel gain from BS $s$ to location $r$. We denote by $\mathcal{P}'= (P_{s}^{(k)})_{1<s<N,\;<k<N_{PRB}}$ the DL Tx traffic channel power matrix. The DL user rate when allocated all resources is given by: 
\begin{equation}
R_D(r,\mathcal{P}') = \sum_{k=1}^{N_{PRB}}W_{PRB}\int^{+\infty}_0 \phi(SD^{(k)}_r\, x) p_{\xi}(x)dx,
\end{equation}

\section{System load and capacity description}\label{sec:flowlevelmetric}
We consider in this Section flow level dynamics, and present the different metrics and key performance indicators (KPIs) used for assessing performance. Consider best effort data traffic with users arriving in the network area at location $r$ according to a Poisson process with rate $\lambda \cdot dr$, where $dr$ represents an infinitessimal surface element. Users upload a file of exponentially distributed size with mean $\mathbb{E}(\sigma_U)< +\infty$ through their serving BS to the network. They are also downloading files of mean size  $\mathbb{E}(\sigma_D)< +\infty$. Relying on the analysis in \cite{bonald2003wireless} on $M/G/1/PS$ queue, we can express the UL (resp. DL) loads of cell $s$, $\bar{\rho}^U_s(\textbf{P})$ (resp. $\bar{\rho}^D_s(\textbf{P})$) and the related KPIs as follow:
\begin{eqnarray}
\bar{\rho}^U_s(\textbf{P}) = \int_{\mathcal{A}_s(\textbf{P})} \frac{\lambda\mathbb{E}(\sigma_U)dr}{R_U(r,\mathcal{P})}\\\nonumber \\ 
C_U(\textbf{P}) = \Big( \int_{\mathcal{A}_s(\textbf{P})}\frac{1}{R_U(r,\mathcal{P})}dr\Big)^{-1} \\\nonumber \\
T_U(r,\textbf{P}) = \frac{\mathbb{E}(\sigma_U)}{R_U(r,\mathcal{P})(1-\bar{\rho}^U_s(\textbf{P}))}\\\nonumber
\end{eqnarray}
Similarly for the DL we have:\\
\begin{eqnarray}
\bar{\rho}^D_s(\textbf{P}) = \int_{\mathcal{A}_s(\textbf{P})} \frac{\lambda\mathbb{E}(\sigma_D)dr}{R_D(r,\mathcal{P}')}\\\nonumber \\
 C_D(\textbf{P}) = \Big( \int_{\mathcal{A}_s(\textbf{P})}\frac{1}{R_D(r,\mathcal{P}')}dr\Big)^{-1}\\\nonumber\\
 T_D(r,\textbf{P}) = \frac{\mathbb{E}(\sigma_D)}{R_D(r,\mathcal{P}')(1-\bar{\rho}^D_s(\textbf{P}))}\\\nonumber
\end{eqnarray}  
where $C_U$ (resp. $C_D$) is the UL (resp. DL) capacity provided by cell $s$ and corresponds to load equals one. $T_U$ (resp. $T_D$) is the UL (resp. DL) file transfer time of user in $r$ associated to cell $s$. 

As discussed previously, the EI defined in the scope of the Lexnet project, requires as input from the network measurements the mean UL Tx and DL Rx powers. The power measurements are aggregated over the periods of users activity, with coefficients which transform power and incident power density into SAR (obtained through measurements and electromagnetic simulations) \cite{EID24lexnet}.
To assess the mean overall UL and DL radiated power in the network, we define the Tx power density $P_{r,s}\lambda(r)dr$ per surface element $dr$. From the power density, we obtain exposure density in term of SAR by the linear transformation $\psi(\cdot)$, namely $\psi(P_{r,s})\lambda(r)dr$. $\psi$ can be regarded as a linear function which weights the measured power with the SAR reference value corresponding to the usage. In compliance to the EI definition, we obtain the mean overall UL exposure level metric as: 
\begin{equation}
f^{UL} = \frac{1}{T}\cdot\frac{\int_{\mathcal{A}} \psi(P_{r,s})T(r) \lambda(r)dr}{|\int_{\mathcal{A}}\lambda(r)dr|},
\end{equation}
and similarly, the mean overall DL exposure level is given by: 
\begin{equation}
f^{DL} = \frac{1}{T}\cdot\frac{C_1\cdot\int_{\mathcal{A}} (\sum_s\psi'(P_{s,r})h_{s,r})T(r) \lambda(r)dr}{|\int_{\mathcal{A}}\lambda(r)dr|},
\end{equation}
where $T(r)$ is the time spent by a user in $r$ in the network and $C_1$ is a factor taking into account exposure during users inactivities. It is estimated in \cite{EID24lexnet} that adult users are active (namely in communication) $5\%$ of the time ($i.e.\; C_1=20$) but are passively exposed to DL EMF the rest of the day.
Note that the discrete versions of expressions of $f^{UL}$ and $f^{DL}$ are nothing but the exposure factor defined in \cite{EID24lexnet}. This factor is the EI term for a specific radio access technology while ignoring the other terms which are not impacted by the self-optimization.

\section{Self-Optimization}\label{sec:sondev}
The optimization under QoS constraint of EMF exposure reduction stated in (\ref{optimpbm}) is addressed in this Section as a self-optimization problem.  
In the following we consider the cell outage as the DL QoS constraint and assume that it is increasing with the load. The latter also increases with the cell coverage and hence with ${\mathbf P^{(0)}} = (P_s^{(0)} + CIO_s)_{ 2<s<N}$, where the macro has the subscript $1$ and is not self-optimized. In the rest of the paper, we interchangeably use ${\mathbf P^{(0)}}$ and ${\mathbf P}$ for simplicity of notations. We define the cell outage at an instant $t$ as the probability that a given user experience a $SINR$ smaller than a predefined threshold $\bar{\theta}$. Equivalently, the outage is the proportion of active users that experience a smaller $SINR$ than this threshold. A closed form expression of the outage probability in mobile networks is proposed in \cite{paris2010outage}. In order to find a solution for (\ref{optimpbm}) and given the DL outage constraint, we design as in \cite{combesInfocom12} an iterative stochastic load-balancing algorithm as follows: 
\begin{equation}\label{iterativeEq}
{\mathbf P^{(0)}}(t+1) = {\mathbf P^{(0)}}(t) + \epsilon \cdot h({\mathbf P^{(0)}}(t))
\end{equation}
where $h({\mathbf P^{(0)}}(t))$ is a vector of length $N-1$ of components $h_s({\mathbf P^{(0)}}(t))$:
\begin{equation}\label{eq:costfunc} h_s({\mathbf P^{(0)}}(t)) =
\left\{
\begin{array}{cc}
\bar{\rho}_M^{UL} - \bar{\rho}_{s}^{UL},& \mbox{if $\theta_s({\mathbf P^{(0)}}(t)) < \bar{\theta}$} \\\\
\bar{\theta} - \theta_s({\mathbf P^{(0)}}(t)) ,& otherwise.
\end{array} 
\right.
\end{equation}
Note that $h_s$ is a discontinuous upper semi-continuous function taking values in the compact convex set $[-1,1]$. The motivation for the definition (\ref{eq:costfunc}) of $h_s$ is to try to balance the macro- and small cells' loads as long as outage is low. Hence when $\theta_s({\mathbf P^{(0)}}(t)) < \bar{\theta}$ the dynamic is driven by the load-balancing objective and the SCs increase their coverages to absorb more UL traffic from the macro station. On the other hand when the condition $\theta_s({\mathbf P^{(0)}}(t)) \geq \bar{\theta}$ holds, the dynamics goes toward the fulfillment of the constraint namely to remain in the constraint set for the outage. We assume that the macro-cell broadcasts regularly its load to the SCs while the outage threshold is introduced by the operator via the management plane. To conduct the convergence analysis of (\ref{iterativeEq}) we need the following proposition: 

\begin{proposition}
The recursive stochastic iteration (\ref{iterativeEq}) is a particular case of the recursive stochastic inclusion 
\begin{equation}\label{recusinclusion}
{\mathbf P^{(0)}}(t+1) = {\mathbf P^{(0)}}(t) + \epsilon \cdot y({\mathbf P^{(0)}}(t)),
\end{equation}
where $y({\mathbf P^{(0)}}(t)) \in g({\mathbf P^{(0)}}(t))$ with $g({\mathbf P^{(0)}}(t)) = \cap_{\gamma>0} \bar{co}(\{h(({\mathbf P'^{(0)}}(t))): ||({\mathbf P'^{(0)}}(t)) - ({\mathbf P^{(0)}}(t))|| < \gamma\}) $ and for a given set $\mathcal{X}$, $\bar{co}(\mathcal{X})$ is the closed convex hull of the set $\mathcal{X}$. \\
Moreover (\ref{recusinclusion}) has the following limiting differential inclusion: 
\begin{equation}\label{recudiffinclusion}
\dot{\mathbf P}^{(0)} \in g({\mathbf P^{(0)}}).
\end{equation}
\end{proposition}

The proof of the proposition is given in Appendix \ref{appendpfprop}.\\
The following theorem discusses the convergence of the iterative dynamic  (\ref{iterativeEq}).
\begin{theorem}
Every solution of the recursive differential inclusion (\ref{recudiffinclusion}) converges to a solution of the original problem (\ref{optimpbm}).\\
\end{theorem}

The proof of the theorem is given in Appendix \ref{appendpfth}.

\section{Numerical analysis}\label{sec:numanalysis}
\begin{figure}[!t]
\centering
\includegraphics[width=3in]{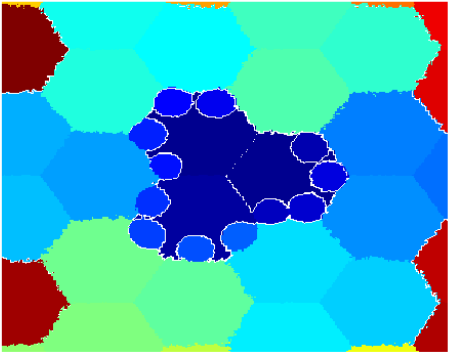}
\caption{Best server Map of the HetNet}
\label{netdeployment}
\vspace{-0.5cm}
\end{figure}

This section presents performance evaluation of the self-optimization algorithm for EMF exposure reduction by means of numerical simulations. 
We consider a geographical area of a LTE network with two rings of tri-sectorized BSs in a dense urban environment. We denote by zone {\bf A} the area covered by the three sectors of the central macro-site and zone {\bf B} the rest of the considered network area. A number of $N_{SCs}$ SCs are deployed in zone {\bf A} close to the cell edge of each macro-cell. Table~\ref{table_simuparam} summarizes the list of the considered parameters and assumptions. In Figure~\ref{netdeployment} we depict the best server map of the considered network with CIO value of 10  set for all SCs.
\begin{table}[!t]
\renewcommand{\arraystretch}{1}
\centering
\begin{tabularx}{\columnwidth}{l|X}
\hline
\textbf{Parameters} & \textbf{Settings} \\
\hline\hline
System configuration & LTE, 10 MHz bandwidth (50 PRBs) UL and DL \\
\hline
Tx powers:      & \\
\hskip0.3cmeNB				     & 46 dBm\\
\hskip0.3cmSCs				    &	 30 dBm. The $CIO$ values are adjusted in the range $-2..10$ dBm \\
\hline
Macro-cell deployment &Hexagonal, 500 m inter-site distance,
57 sectors simulated, statistics maintained over 3
central sectors\\
\hline
SC deployment & 4 SCs per sectors located randomly closed to the egde of each sector coverage\\
\hline
User deployment & Poisson Arrival at rate $\lambda =5$. Maximum of $8$ users in other cells\\
\hline
Service Type& File download/Upload\\
File Size& 15Mbits\\
UL/DL traffic load ratio& 30\%\\
Coverage target rate & 1.5Mbps\\
\hline
Pathloss & \\
\hskip0.3cmMacro-to-UE& $128.1+37.6\log10(R)$\\
\hskip0.3cmSCs-to-UE& $140.7+36.7\log10(R)$\\
Shadowing standard deviation& 6 dB\\
\hline
Scheduler & Round Robin\\
\hline
Exposure&\\
\hskip0.3cm DL SAR weight& 4.7e-3 W/Kg\\
\hskip0.3cm UL SAR weight& 8e-5 W/Kg\\
\hskip0.3cm Users activity coef. & 20\\
\hline

\end{tabularx}
\caption{Simulation parameters settings}\label{table_simuparam}
\end{table}
\paragraph{Uplink interference structure}
To minimize truncation effects of the network simulation area, and to obtain realistic performance in zone {\bf A}, 
it is important to take into account interference generated at zone {\bf B}. We assumed that the traffic in the cells of zone {\bf B} is strongly correlated to that of the cells in zone {\bf A}, with a maximum of eight users per cell.
%

\begin{figure}[!t]
\centering
\includegraphics[width=3in]{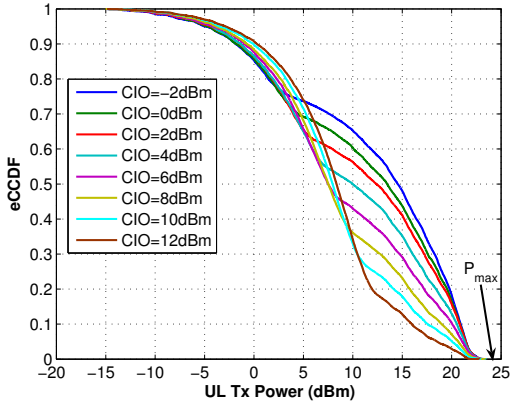}
\caption{UL Tx Power for increasing values of $CIO$}
\label{ultxpwinccio}
\vspace{-0.5cm}
\end{figure}

\begin{figure}[!t]
\centering
\includegraphics[width=3in]{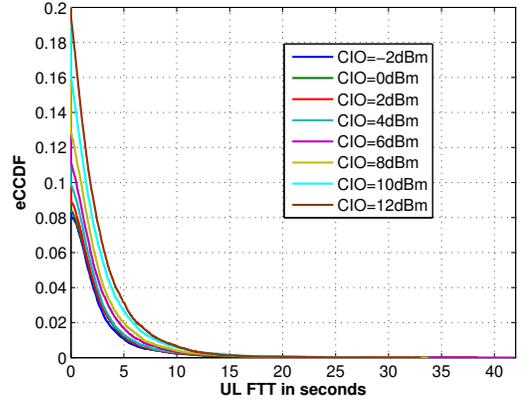}
\caption{UL File Transfer Time for increasing values of $CIO$}
\label{ulfttinccio}
\vspace{-0.5cm}
\end{figure}

\paragraph{ABS mute ratio}
Refererence \cite{tall14wiopt} shows that the performance achieved using load-balancing or the Cell Range Extension (CRE) can be significantly improved when applied in conjunction with  Almost Blank Sub-frame (ABS) self-optimization.  The latter reduces macro-cell interference produced on users at the extended coverage area of the SCs by muting a portion of the macro station transmissions. This mechanisms is adopted in the present work.

\subsection{Performance bounds}
In an initial set of simulations, we increase SCs coverage using the CIO. The objective here is to highlight the relation between the CIO parameter and QoS, overall radiate power and EI. In Figures~\ref{uldlmeanexpoincio}-\ref{dlmeanoutageinccio} and \ref{sonoutage}, the upper 3 curves corresponds to the macro-cell while the lower ones - to the SCs.
\subsubsection{Base line setting}
 We start with the baseline setting when SCs do not expand their coverage. In this setting, we observe that most of the arriving users are connected to the macro-cell since it offers a better received signal. However in the UL, due to high path-losses, many macro-cells users transmit with a high power. In Figure~\ref{ultxpwinccio} we plot the empirical complementary cumulative distribution of the users UL Tx power. The upper curve represents the distribution of power for the first applied value of $CIO = -2dBm$. This is indeed the case where the overall UL Tx power is the highest which results in the highest level of the exposure. Note that for each curve (in Figure~\ref{ultxpwinccio}) the upper part is representative of a lower Tx power from the SCs users compared to macro users Tx power, which contribute more in the lower part. 

\subsubsection{Impact of coverage expansion}
\begin{figure}[!t]
\centering
\includegraphics[width=3in]{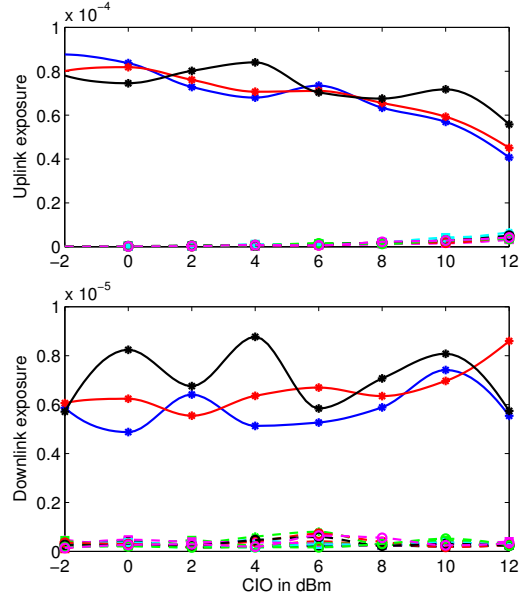}
\caption{UL and DL mean exposure for increasing values of $CIO$ for the macro cells (higher three curves) and the small cells (lower curves)}
\label{uldlmeanexpoincio}
\vspace{-0.5cm}
\end{figure}

Now we start increasing the coverage of SCs. When SCs coverages are small, there are very few users that are off-loaded while the macro remains highly loaded (the scenario considers a high load regime). As a consequence, the cell outage is very high at the macro while it remains low at the SCs. However as the SCs coverage expands, their users experience growing interference. In parallel, with the increase of CIO, SC users in the extended coverage area will increase their Tx power. Due to these two combined factors, the UL FTT at the SCs thus increases as shown in Figure~\ref{ulfttinccio}. Interestingly this increase remains low enough to observe a good trade-off compared to the important decrease in UL Tx power. This is what we observe in Figure~\ref{uldlmeanexpoincio} for the UL exposure which decreases despite the light interference impairments. One can also see that the DL has a limited contribution to overall exposure. Again as we noted in Figure~\ref{ultxpwinccio}, the macro users UL Tx power is decreasing since they are being closer to their serving cells. As a result SCs users perceive less interference from the macro users which contributes to the aforementioned interference impairments.

\begin{figure}[!t]
\centering
\includegraphics[width=3in]{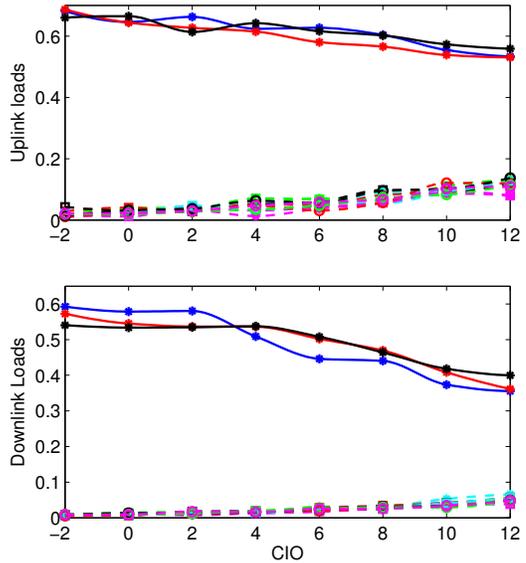}
\caption{UL and DL loads for increasing values of $CIO$}
\label{uldlloadinccio}
\vspace{-0.5cm}
\end{figure}

\begin{figure}[!t]
\centering
\includegraphics[width=3in]{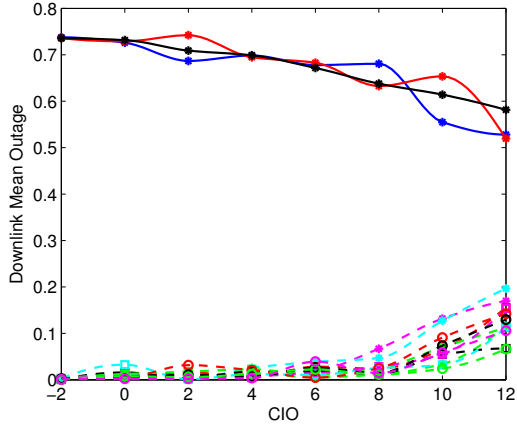}
\caption{DL mean outage for increasing values of $CIO$}
\label{dlmeanoutageinccio}
\vspace{-0.5cm}
\end{figure}

The evolution of the load as a function of CIO is shown in Figure~\ref{uldlloadinccio}. In the DL users accumulate at the SCs nodes while they depart fast from the macro as they are fewer, and are served with fixed DL power. One can see that the macro-cell load decreases with the increase in CIO, and remain significantly higher than the SCs' loads.
A similar trend is observed in Figure~\ref{dlmeanoutageinccio} for the outage. One can see that the mean outage decreases at the macro-cell and grows at SCs nodes for high CIO values until they are overloaded and can barely maintain an acceptable QoS. It is noted that not all SCs benefit from coverage increase, thus motivating the use of a self-optimization.  


\subsection{Self-optimized scenario}
\begin{figure}[!t]
\centering
\includegraphics[width=3in]{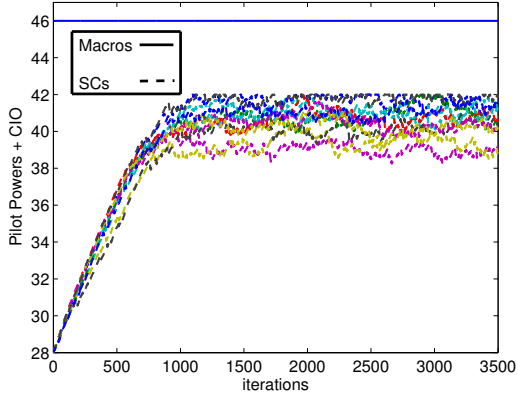}
\caption{Convergence of $Pilote Powers +CIO$ values using self-optimization}
\label{pilotecioadpt}
\vspace{-0.5cm}
\end{figure}

\begin{figure}[!t]
\centering
\includegraphics[width=3in]{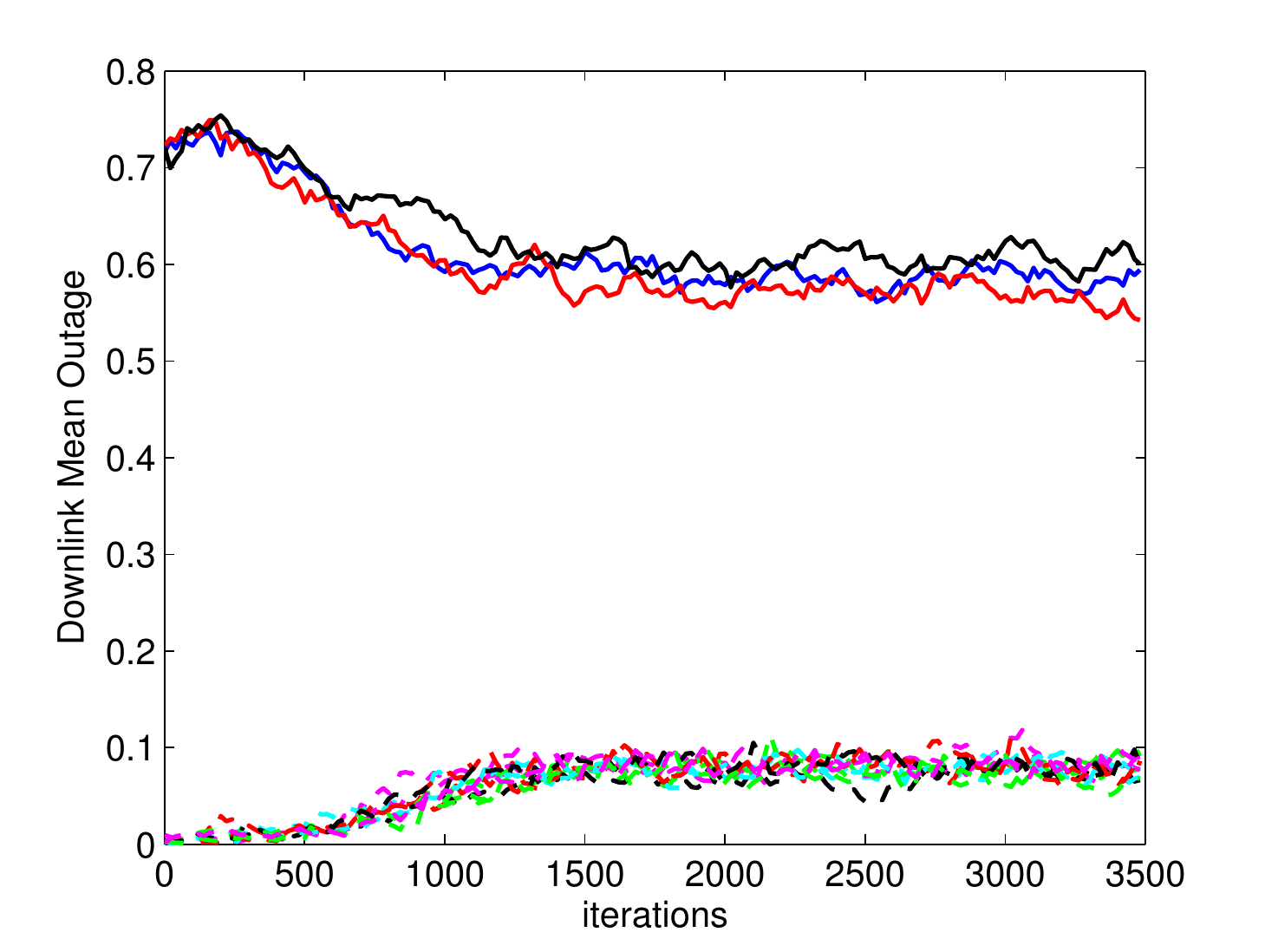}
\caption{DL mean outage values using self-optimization}
\label{sonoutage}
\vspace{-0.7cm}
\end{figure}

Under the same setting as before we now activate the distributed self-optimizing algorithm in the SCs with a target outage of $5\%$ and a threshold rate of $1.5Mbps$. In Figure~\ref{pilotecioadpt} we plot the evolution of pilot power plus $CIO$ as a function of the algorithm iterations (6s per iteration). We observe that the $CIO$ does not systematically grow to its upper limit of $12 dBm$ and adapts to the QoS constraint according to the traffic in each SC. Note that the pilot powers from the macro-cells are kept constant as they are not concerned with the coverage expansion. In Figure~\ref{sonoutage} we can see that the outage barely grows above the defined limit and is indeed monitored and controlled by the autonomous mechanism. 

\begin{figure}[!t]
\centering
\includegraphics[width=3in]{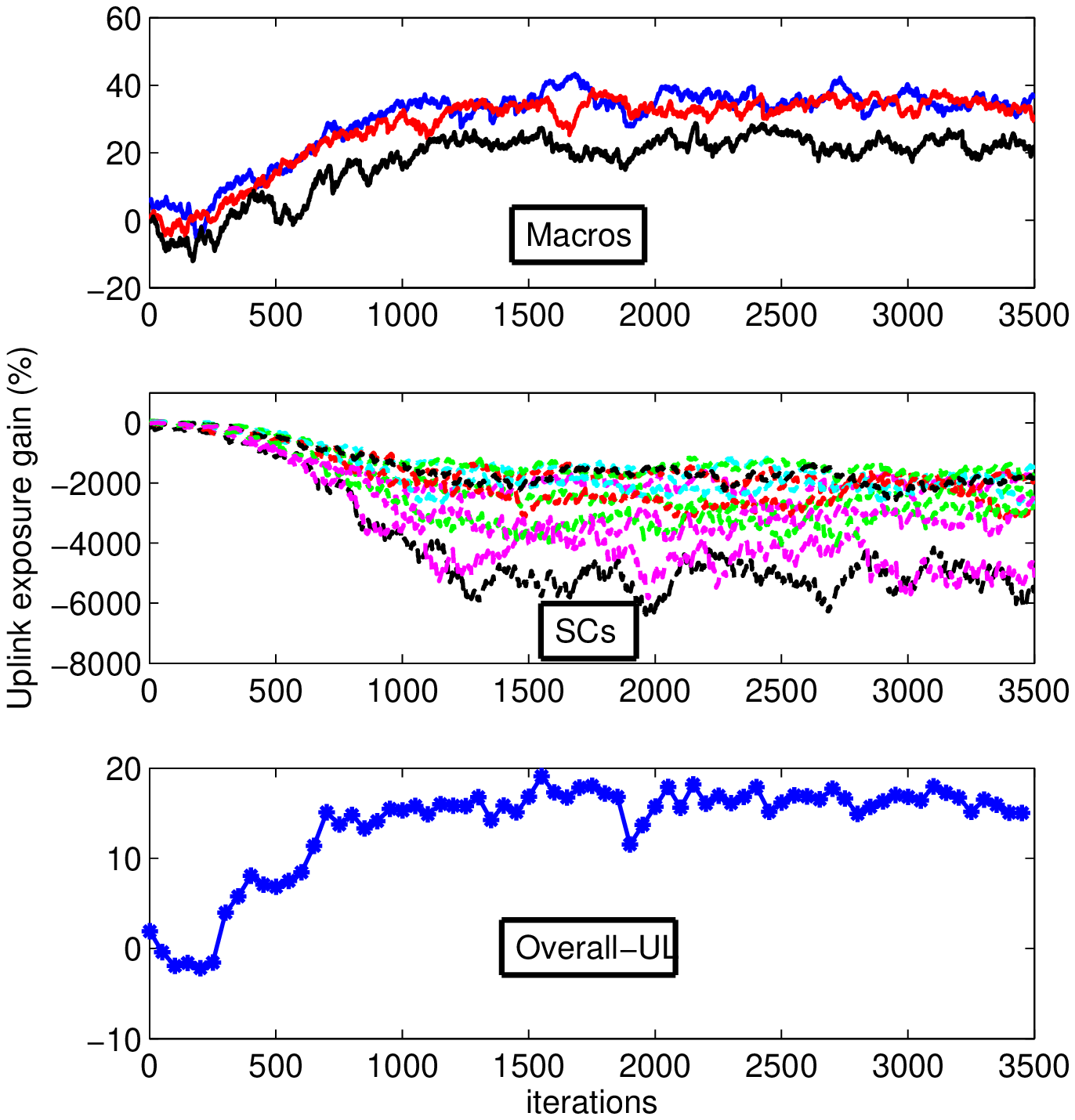}
\caption{Percentage of mean UL exposure gain using self-optimization}
\label{gainulexposure}
\vspace{-0.7cm}
\end{figure}

\begin{figure}[!t]
\centering
\includegraphics[width=3in]{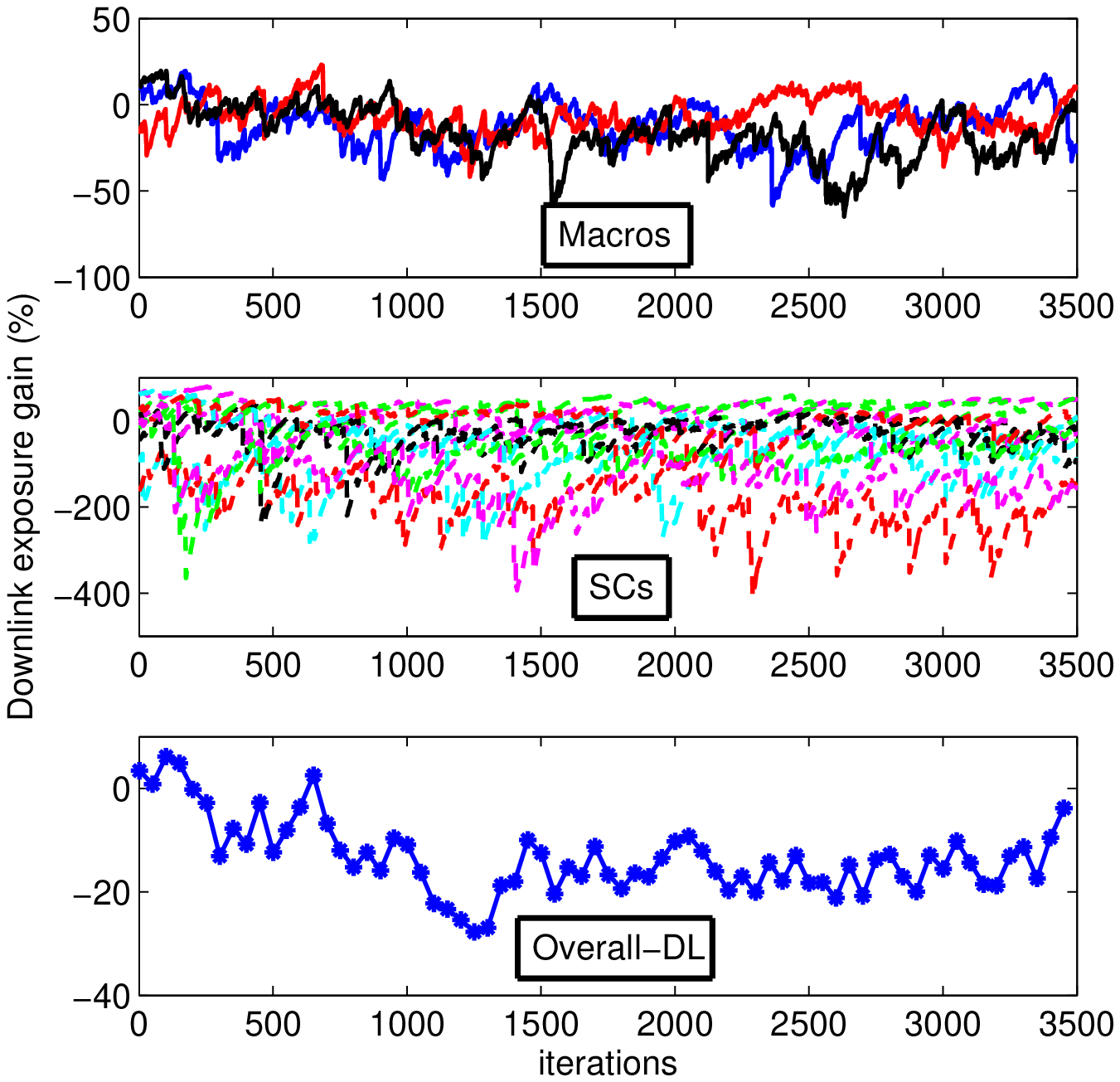}
\caption{Percentage of mean DL exposure gain using self-optimization}
\label{gaindlexposure}
\vspace{-0.5cm}
\end{figure}

\begin{figure}[!t]
\centering
\includegraphics[width=3in]{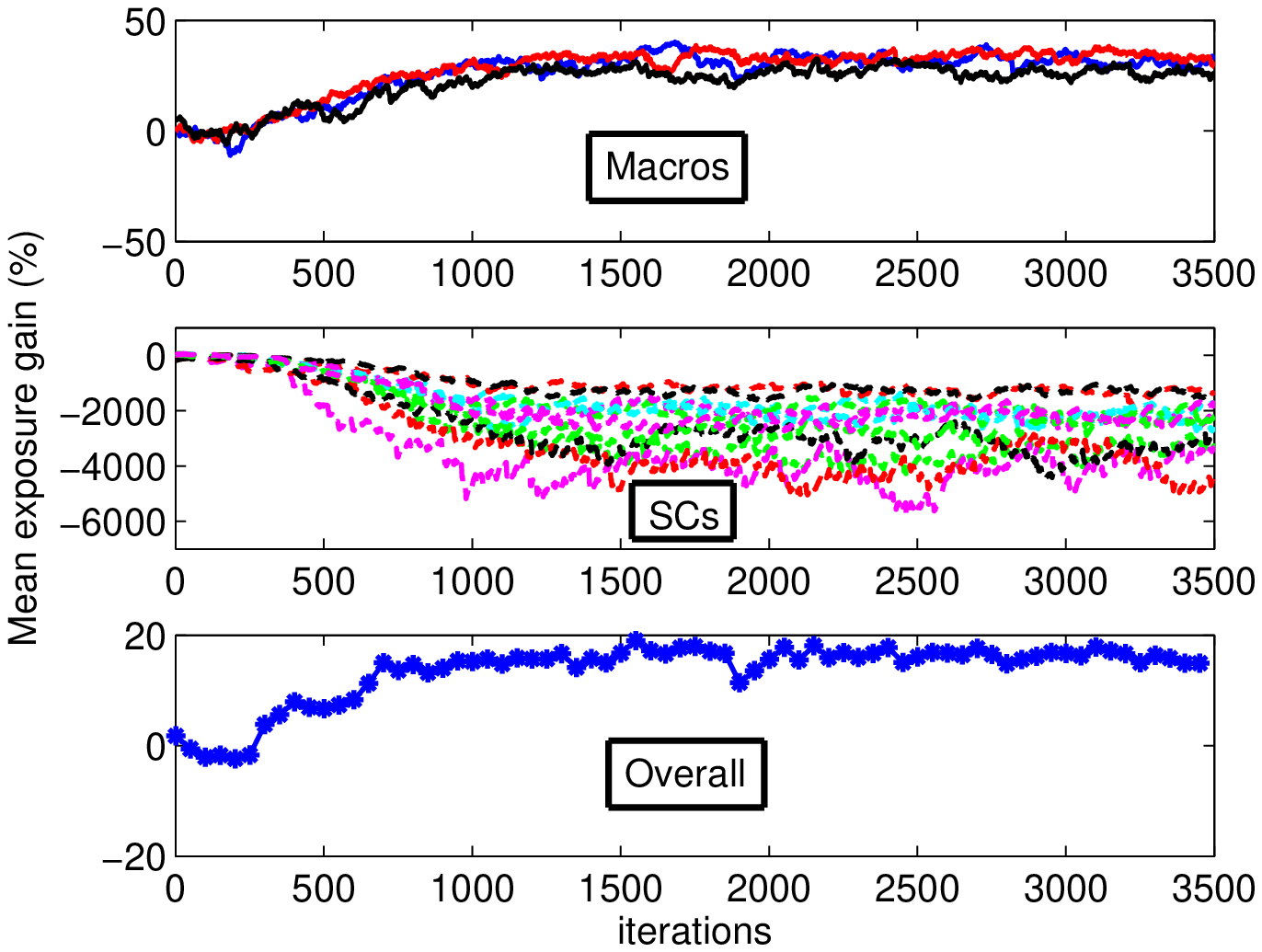}
\caption{Percentage of mean EMF exposure gain for combined UL and DL using self-optimization}
\label{gainmeanexposure}
\vspace{-0.5cm}
\end{figure}

We define the exposure gain as the relative exposure reduction with respect to the base line scenario. A reduction in exposure is translated into a positive gain, and vise versa.    
Figures~\ref{gainulexposure}, \ref{gaindlexposure} and \ref{gainmeanexposure} show the main result of the numerical analysis, namely the exposure gain for the UL, DL and the total (i.e. UL and DL combined) EI, respectively, brought about by the self-optimizing algorithm. 
The combined UL and DL average exposure gain for the macro-cells is around $30$ percent (see Figure \ref{gainmeanexposure}), which is similar to that in the UL (see Figure  \ref{gainulexposure}). In the DL, a relative exposure increase varying between 10 and 20 percent can be observed (see lower curve in Figure \ref{gaindlexposure}), which is accompanied by an increasing outage of DL users at the SCs. It is recalled that the outage level is controlled by the self-optimizing algorithm. The overall EMF reduction computed over the cells implementing the SON algorithm varies between 15 and 20 percent.


\section{Conclusion}\label{sec:conclusion}
With the objective of reducing the overall exposure to EMF in mobile networks this work has addressed the assessment and reduction of RF exposure relying on macro-cells off-loading with SCs nodes. Using the definition of the EI provided by the European FP7 Lexnet project, which considers the combined effect of UL and DL transmissions, we have modeled the problem as a distributed optimization of UL KPIs under the constraint of DL QoS. This is one of the main output of this work where we show and utilize the strong correlation between UL performances improvement and DL QoS degradation when dealing with the reduction of the EI. We have designed a controlled stochastic load-balancing mechanism to perform an iterative reduction of the defined exposure cost function for which the algorithm has been proved to converge. The results of this study lead naturally to several
 directions of future investigations. In fact the EI as defined does not comply with on-line monitoring in order to be assessed. This opens a wide panel of candidate solutions relying on the principal levers impacting EMF exposure, as considered in this paper. Finally encouraging results have been obtained after convergence of the proposed mechanism which achieves an overall exposure gain varying between 15 and 20 percents. 

\section*{Acknowledgment}
This paper reports work undertaken in the context of the project LEXNET. LEXNET is a project supported by the European Commission in the 7th Framework Program (GA $n^\circ318273$). For further information, please visit \mbox{\url{www.lexnet-project.eu}}.

The authors would like to thank Dr. Richard Combes, Dr. Emmanuelle Conil and Dr. Joe Wiart for the enriched discussions that have contributed to the completion of this work.

\appendices
\section{Proof of proposition 1}\label{appendpfprop}
\begin{proof}
The first part of the proof is straight forward. We have for all component $g_s({\mathbf P^{(0)}}(t))$ of the vector $g({\mathbf P^{(0)}}(t))$ : $g_s({\mathbf P^{(0)}}(t)) =$\\
$$\cap_{\gamma>0} \bar{co}(\{h_s(({\mathbf P'^{(0)}}(t))) : ||({\mathbf P'^{(0)}}(t)) - ({\mathbf P^{(0)}}(t))|| < \gamma\}). $$ 
$\forall \; {\mathbf P^{(0)}}(t) \; s.t. \; \theta_s({\mathbf P^{(0)}}(t)) < \bar{\theta}\;or\;\theta_s({\mathbf P^{(0)}}(t)) > \bar{\theta},\; h_s({\mathbf P^{(0)}}(t)) = g_s({\mathbf P^{(0)}}(t))$. \\
When $\theta_s({\mathbf P^{(0)}}(t)) = \bar{\theta}$ then : 
\begin{eqnarray}
g_s({\mathbf P^{(0)}}(t)) = \bar{co}(\{ h_s(({\mathbf P^{(0)}}(t))),\, \bar{\rho}_1^{U} - \bar{\rho}_{s}^{U} \})\\ \nonumber
\mbox{thus, } h_s(({\mathbf P^{(0)}}(t))) \in g_s({\mathbf P^{(0)}}(t))  \nonumber
\end{eqnarray}
which proves that (\ref{iterativeEq}) is a particular case of (\ref{recusinclusion}). \\
To prove that (\ref{recudiffinclusion}) is indeed the limiting differential inclusion for (\ref{recusinclusion}), based on the results in \cite[Lemma 1, p. 52-59]{borkar2008stochastic}  we need the function $g$ to fulfill the following conditions which are verified here. 
\begin{itemize}
\item[{\it (i)}] $g({\mathbf P^{(0)}}(t))$ is convex and compact for each ${\mathbf P^{(0)}}(t)$. By construction of $g$ and as $h$ takes value in the compact set $[-1,1]^{N-1}$ this is verified here.
\item[{\it (ii)}] $\forall \;{\mathbf P^{(0)}}(t)$ it must exist some $K>0$ such that : $\underset{y\in g({\mathbf P^{(0)}}(t))}{\sup}||y||< K(1 + ||{\mathbf P^{(0)}}(t))||)$. We have:\\
 \small
\[\hskip-1cm\underset{y\in g({\mathbf P})}{\sup}||y||=
\left\{
\begin{array}{cc}
||\bar{\rho}_1^{U} - \bar{\rho}_{s}^{U}||,& \mbox{if $\theta_s({\mathbf P}) < \bar{\theta}$} \\\\
||\bar{\theta} - \theta_s({\mathbf P^{(0)}}(t))|| ,& \mbox{if $\theta_s({\mathbf P}) > \bar{\theta}$} \\\\
\sup\{||u(\bar{\rho}_1^{U} - \bar{\rho}_{s}^{U}) +& \\
\hskip-0.2cm(1-u)(\bar{\theta} - \theta_s({\mathbf P}))||,\,u\in[0,1]\},& otherwise
\end{array}
\right.
\]\normalsize
For the two first cases the function $\underset{y\in g({\mathbf P})}{\sup}||y||$ is pseudo-contractive as it shows as a particular form of the discussed function in \cite[Theorem 3]{combesInfocom12}. In the discussions over there both terms of the difference function change with the parameter whereas in our case one term is fixed. The function in the third case is a linear combination of the functions of the first two cases which are pseudo-contractive. The result follows as a consequence. 
\item[{\it (iii)}] $g$ is upper semi-continuous, which is insured by the upper semi-continuity of $h$.
\end{itemize}
This completes the proof.
\end{proof}

\section{Proof of theorem 1}\label{appendpfth}
\begin{proof}
From \cite[Corollary 4, p. 55]{borkar2008stochastic} it is sufficient, given the properties of $g$ from proposition 1, to show that : 
$$
\underset{t}{sup}||{\mathbf P^{(0)}}(t)||<\infty
$$
This is particularly true when there exists a Lyapunov function $V(\cdot)$ of the dynamic induced by $g$ verifying \cite[Lemma 1 and Theorem 2 p. 12-15]{borkar2008stochastic}: 
\begin{itemize}
\item[{\it (i)}] $\underset{||{\mathbf P^{(0)}}||\rightarrow \infty}{lim} V({\mathbf P^{(0)}})=\infty$\\
\item[{\it (ii)}] $\dot{V} \leq 0$ and $\dot{V} < 0$ outside the bounded set of the constrains.
\end{itemize}
Let, $$V({\mathbf P}) = \underset{s}{\max}(\mathbbm{1}_{\{\theta_s>\bar{\theta}\}}||{\mathbf P}||(\theta_s({\mathbf P}) - \bar{\theta})^2))$$
 we have $\underset{||{\mathbf P}||\rightarrow \infty}{lim} V({\mathbf P})=\infty$. Moreover by restricting  to the case $\theta_s>\bar{\theta}$ and making the derivative along $s_{max} = arg \underset{s}{\max}(\mathbbm{1}_{\{\theta_s>\bar{\theta}\}}||{\mathbf P}||(\theta_s({\mathbf P}) - \bar{\theta})^2))\; \forall \; {\mathbf P}$, we have:
\begin{eqnarray*}
\dot{V}&=& 2||{\mathbf P}||\dot{\theta}_{s_{max}({\mathbf P})}(\theta_{s_{max}}({\mathbf P}) - \bar{\theta}) \\\nonumber
&=& 2||{\mathbf P}||\frac{\delta \theta_{s_{max}}({\mathbf P})}{\delta{\mathbf P}}\frac{\delta {\mathbf P}}{\delta t}\vert_{\{\theta_s>\bar{\theta}\}}(\theta_{s_{max}}({\mathbf P}) - \bar{\theta})\\\nonumber
&=& -2||{\mathbf P}||\frac{\delta \theta_{s_{max}}({\mathbf P})}{\delta{\mathbf P}}(\theta_{s_{max}}({\mathbf P}) - \bar{\theta})^2\nonumber
\end{eqnarray*}
As $\theta_s$ is increasing with ${\mathbf P},\; \forall\; s$ we have $\dot{V}\leq0$ which concludes the proof.
\end{proof}


\end{document}